\documentclass[a4paper]{jpconf}

\usepackage{bm}

\begin{document}

\title{{Methods for Derivation of Generalized Equations 
in the $(S,0)\oplus(0,S)$ Representations of the Lorentz Group}\footnote{Some parts of this paper have been presented at the XI Escuela de DGFM SMF, Dec. 5-9, 2016, Playa del Carmen, QRoo, M\'exico, the IARD2018, June 4-7, 2018, M\'erida, Yuc., M\'exico
and the MG15 Meeting, July 1-7, Rome, Italy. To be published in the Proceedings of the 11th Vigier Symposium, Aug. 6-9, 2018. Liege, Belgium.}}

\author{Valeriy V. Dvoeglazov}

\address{UAF, Universidad Aut\'onoma de Zacatecas\\
Apartado Postal 636, Suc. 3, Zacatecas 98061 Zac., M\'exico\\
E-mail: valeri@fisica.uaz.edu.mx}

\date{\empty}


\begin{abstract}
We continue the discussion of several explicit examples of generalizations in relativistic quantum mechanics.
We discussed the generalized spin-1/2 equations for neutrinos and the spin-1 equations for photon. The equations  obtained by means of the Gersten-Sakurai method and those of Weinberg for spin-1 particles  have been mentioned. Thus, we generalized the Maxwell and Weyl equations. Particularly, we found connections of the well-known solutions and the dark 4-spinors in the Ahluwalia-Grumiller elko model. They are also not the eigenstates of the chirality and helicity. 
The equations may lead to the dynamics which are different from those accepted at the present time. For instance, the photon may have non-transverse components and the neutrino may be {\it not} in the energy states and in the chirality states.
The second-order equations have been considered too. They have been obtained by the Ryder method.
\end{abstract}

\section{Introduction. The Gersten Method and its Relation to Relativistic Quantum Equations.}

The content of this talk is the following. We use
the van der Waerden-Sakurai procedure for derivation of the Dirac equation and its generalizations. As a consequence, 
for example, the scalar fields appear in the Maxwell-like equations. Next, we have 
massless particles ($p^2 =0$) with massive parameters, which may be considered as the measures of chirality.
We also discuss the massive/massless 2nd order equations.

The first part is based on recent papers~\cite{Ger,dvo1,dvon}.
Gersten~[1a] writes: ``We have shown how all Maxwell equations can be
derived simultaneously from first principles, similar to those which have
been used to derive the Dirac relativistic electron equation" and
concludes: ``\ldots Maxwell equations should be used as a guideline for
proper interpretation of quantum theories".

In fact, he used a method presented by van der Waerden and
Sakurai~\cite{Sak}. Let us begin with the Klein-Gordon equation:
\begin{equation}
( E^2 -c^2 {\bf p}^2 -m^2 c^4 )  \Psi_{(2)} =0\,.\label{kg}
\end{equation}
Hence, for $\Psi$ with {\it two} components and
$c=\hbar=1$ we have
\begin{equation}
(E I^{(2)}-{\bm \sigma}\cdot {\bf p}) (E I^{(2)}+ {\bm\sigma}\cdot
{\bf p} ) \Psi_{(2)} = m^2 \Psi_{(2)} \,.
\end{equation}
Let us denote
$\Psi_{(2)} =\eta$, then we have
  \begin{eqnarray} {(E
I^{(2)}+{\bm \sigma}\cdot {\bf p})\over m} \eta &=&\chi\,,\\
(E I^{(2)} -
{\bm\sigma}\cdot {\bf p} ) \chi &=& m \eta \,.  \end{eqnarray}
In the 4-component form one has
\begin{eqnarray}
\pmatrix{-m I^{(2)}& E I^{(2)}+{\bm \sigma}\cdot {\bf p}\cr
E I^{(2)}-
{\bm\sigma}\cdot {\bf p}& -mI^{(2)}\cr} \pmatrix{\chi\cr\eta\cr} =0
\end{eqnarray} With the quantum-operator substitutions $E\rightarrow
i\hbar {\partial \over \partial t}$ and ${\bf p} \rightarrow -i\hbar
{\bm\nabla}$ we recover the Dirac equation!\footnote{One can also
decompose (\ref{kg}) into the 4-component form from the beginning $$(E
I^{(4)}+{\bm \alpha}\cdot {\bf p} +m\beta) (E I^{(4)}-{\bm\alpha}\cdot
{\bf p} -m\beta ) \Psi_{(4)} =0$$ and then look for ${\bm\alpha}$ and
$\beta$.}

In the $S=1$ and $m=0$ case one can proceed in a similar way:
\begin{eqnarray}
\left ( {E^2\over c^2}   - {\bf p}^{\,2} \right ) {\bm \Psi}_{(3)}
&=& \left ( {E\over c} I^{(3)} - {\bf S}\cdot {\bf p} \right )
\left ( {E\over c} I^{(3)} + {\bf S}\cdot {\bf p} \right )
{\bm \Psi}_{(3)} - \label{Gersten1}\\
&-&\pmatrix{p_x\cr p_y \cr p_z\cr}
\left ({\bf p}\cdot {\bm \Psi}_{(3)}\right ) =0
\,,\quad \mbox{Eq. (9) of~[1a]}\,,
\nonumber
\end{eqnarray}
where $({\bf S}^i)^{jk} = -i\epsilon^{ijk}$.\footnote{Please note
the difference $({\bm\sigma}\cdot {\bf p})^2 ={\bf p}^{\,2}$,
but $({\bf S}\cdot {\bf p})^3 = ({\bf S}\cdot {\bf p})$; the spin-1
matrices are {\it singular}.}
Gersten found that solutions are defined by 
\begin{eqnarray}
&&\left ( {E\over c} I^{(3)} +{\bf S}\cdot {\bf p} \right )
{\bm\Psi}_{(3)} =0\quad\qquad\mbox{Eq. (10) of~[1a]}\nonumber\\
&&\left ({\bf p}\cdot {\bm \Psi}_{(3)}\right ) =0
\qquad\qquad\mbox{Eq. (11) of~[1a]}\nonumber
\end{eqnarray}
and their complex conjugates. The latter may be interpreted
as the solutions of opposite helicity.
If one assumes ${\bm\Psi}_{(3)} = {\bf E} - i{\bf B}$ then after quantum
operator substitutions: 
\begin{eqnarray}
&&{i\hbar\over c} {\partial {\bm \Psi}_{(3)} \over \partial t}
= -\hbar {\bm\nabla}\times {\bm\Psi}_{(3)}\,,\\
&&-i\hbar {\bm \nabla}\cdot {\bm\Psi}_{(3)} =0\,,
\end{eqnarray}
or
\begin{eqnarray}
&&{\bm\nabla}\times ({\bf E}-i{\bf B}) = -{i\over c} {\partial ({\bf
E}-i{\bf B}) \over \partial t}\,,\\
&& {\bm\nabla} \cdot ({\bf E} - i{\bf B}) =0\,.
\end{eqnarray}  
Please note that the Planck constant was cancelled out!
The procedure of separation  into real and imaginary parts leads to
\begin{equation}
{\bm \nabla}\times {\bf
\vec{E}}=-\frac{1}{c}\frac{\partial {\bf \vec{B}}}{\partial t} \,,
\end{equation}
\begin{equation}
{\bm \nabla }\times {\bf \vec{B}}=\frac{1}{c}\frac{\partial {\bf
\vec{E}}}{\partial t} \,,
\end{equation}
\begin{equation}
{\bm \nabla }\cdot {\bf \vec{E}}= 0\,,
\end{equation}
\begin{equation}
{\bm \nabla }\cdot {\bf \vec{B}}=0\,.
\end{equation}
The situation is the same if one starts with the complex conjugate function
${\bm\Psi}^\ast_{(3)} = {\bf E}+i{\bf B}$.

The Lagrangian for this theory has recently been
presented in~\cite{Ger,Hott}:
\begin{equation}
{\cal L} = -c {\bm\Psi}^\dagger_{(3)} ({E\over c} I^{(3)}
+{\bf S}\cdot {\bf p} )\ {\bm \Psi}_{(3)}\,.
\end{equation}
However, it was shown in~\cite{dvoo} that such a form of Lagrangians is
{\it not} a scalar (it is a zero component of a 4-vector; see
also~\cite{sud}). Next, we have
\begin{equation}
W_\mu W^\mu = -s(s+1) p_\mu p^\mu I^{(s)}
\end{equation}
and
\begin{equation}
W^\mu p_\mu =0\,
\end{equation}
by definition. Hence,
\begin{equation}
(W_\mu -s p_\mu) (W^\mu +sp^\mu ) =-s (2s+1) p_\mu p^\mu I^{(s)}\,,
\end{equation}
which is similar to Eq. (9) of Ref.~[1a] if one imposes $p_\mu p^\mu
=m^2=0$.

Therefore, on using the definition of the Pauli-Lubanski operator we have: 
\begin{eqnarray}
&&({\bf S}\cdot {\bf p} - s p_0 ) {\bm\Psi}_{(s)}=0\,,\\
&&({\bf S} p_0 +i {\bf S}\times {\bf p} - s {\bf p}) {\bm \Psi}_{(s)}
=0\,.  \end{eqnarray}  
It is easy to see that in the
$S=1$ case one can recover the previous consideration~[1b]. The second
equation can be considered as a subsidiary condition
\begin{eqnarray}
\pmatrix{0&0&0\cr p_x&p_y&p_z\cr 0&0&0\cr} {\bm\Psi}_{(3)} =0\,.
\end{eqnarray}
In~\cite{dvo1} we corrected the Gersten's claim. The equation (9) of Ref.~[1a]
(the equation (\ref{Gersten1}) above) is satisfied also under the choice 
\begin{eqnarray}
&&\left ( {E\over c} I^{(3)} +  {\bf S}\cdot {\bf p} \right )
{\bm\Psi}_{(3)} ={\bf p} \chi\,,\\
&&\left ({\bf p}\cdot {\bm \Psi}_{(3)}\right ) ={E\over c} \chi\, ,
\end{eqnarray}
due to $({\bf S}\cdot {\bf p}) {\bf p} =0$. The above set leads to
 \begin{eqnarray}
&&{\bm \nabla}\times {\bf
E}=-\frac{1}{c}\frac{\partial {\bf B}}{\partial t} + {\bm
\nabla} {\it Im} \chi \,, \label{1}\\
&&{\bm \nabla }\times {\bf B}=\frac{1}{c}\frac{\partial {\bf
E}}{\partial t}  +{\bm \nabla} {\it Re} \chi\,,\label{2}\\
&&{\bm \nabla}\cdot {\bf E}=-{1\over c} {\partial \over \partial
t} {\it Re}\chi \,,\label{3}\\
&&{\bm \nabla }\cdot {\bf B}= {1\over
c} {\partial \over \partial t} {\it Im} \chi \,,  \label{4}
\end{eqnarray}
with an additional scalar field $\chi$.
It is also possible that 
\begin{eqnarray}
&&\left ( {E\over c} I^{(3)} - {\bf S}\cdot {\bf p} \right )
{\bm\Psi}_{(3)}^\prime ={\bf p} \chi^\prime\\
&&\left ({\bf p}\cdot {\bm \Psi}^\prime_{(3)}\right ) ={E\over c}
\chi^\prime\, .
\end{eqnarray} 
If $\chi^\prime = \chi^\ast$ then ${\bm \Psi}^\prime_{(3)} = {\bf
E}+i {\bf B}$ and we recover the generalized Maxwell equations
(\ref{1}-\ref{4}). Kruglov found relations of $\chi -$ functions with QED and the Riemann 
tensor~\cite{Kruglov}.

Thus,
\begin{itemize}
\item
We obtain ${\bf p}\cdot {\bm \Psi}_{(3)} \neq 0$, therefore the free
photon may have a non-transverse component~\cite{ima};

\item
The $\chi$-fields may be function(al) of higher-rank tensor fields, thus
leading to equations which are non-linear in ${\bf E}$ and ${\bf B}$
(cf. Ref.~\cite{donev});

\item
One can find possible relations to the
Ogievetski\u{\i}-Polubarinov-Kalb-Ramond field~\cite{op,kr,dvob}. After
performing the Bargmann-Wigner procedure for the spin-1 field we obtain 
\begin{eqnarray}
&&\partial_\alpha F^{\alpha\mu} +{m\over 2} A^\mu =0\,,\label{p1}\\
&&2mF^{\mu\nu} = \partial^\mu A^\nu -\partial^\nu A^\mu\,,\label{p2}
\end{eqnarray} 
instead of the well-known Proca set: 
\begin{eqnarray}
&&\partial_\alpha F^{\alpha\mu} +m^2 A^\mu =0\,,\\
&&F^{\mu\nu} = \partial^\mu A^\nu -\partial^\nu A^\mu\,,
\end{eqnarray}
 In fact, these sets are related  one to another by the re-normalization
transformation:  $A^\mu \rightarrow 2mA^\mu$ or $F^{\mu\nu} \rightarrow
{1\over 2m} F^{\mu\nu}$. Ogievetski\u{\i} and Polubarinov~\cite{op} wrote:
``In the massless limit the system of $2s+1$ states is no longer
irreducible;
it decomposes and describes a set of different particles with zero mass
and helicities $\pm s,\pm (s-1), \pm 1, 0$ (for integer spin and if parity
is conserved); the situation is analogous for half-integer
spins." One can also see this after performing the Lorentz
transformation in the $(1/2,1/2)$ representation, $A^\mu =
\Lambda^\mu_{\quad\nu} A^\nu_{basis}$, and looking for massless limit.
Thus, we find
\begin{equation}
u^\mu
({\bf p}, +1)= -{N\over \sqrt{2}m}\pmatrix{-p_r\cr m+ {p_x p_r \over
E_p+m}\cr im +{p_y p_r \over E_p+m}\cr {p_z p_r \over
E_p+m}\cr}\,,\quad  u^\mu ({\bf p}, -1)= {N\over
\sqrt{2}m}\pmatrix{-p_l\cr m+ {p_x p_l \over E_p+m}\cr -im +{p_y p_l \over
E_p+m}\cr {p_z p_l \over E_p+m}\cr}\,,\label{vp12}
\end{equation}
and
\begin{equation}
u^\mu ({\bf
p}, 0) = {N\over m}\pmatrix{-p_z\cr {p_x p_z \over E_p+m}\cr {p_y p_z
\over E_p+m}\cr m + {p_z^2 \over E_p+m}\cr}\,,\quad
u^\mu ({\bf p}, 0_t) = {N \over m} \pmatrix{E_p\cr -p_x
\cr -p_y\cr -p_z\cr}\,.\label{vp3}
\end{equation}
Please note that for helicities $\sigma = \pm 1, 0$
one has $p_\mu u^\mu ({\bf p}, \sigma ) = 0$ (an analogue of the
Lorentz condition~\cite{dvob}). This is
not the case for the ``time-like" photons. In view of the fact that in the case
$N=1$ we have divergent
behaviour of  certain parts of the 4-vector momentum-space functions in
$m\rightarrow 0$, the first degree of $m$ in the equations
(\ref{p1},\ref{p2}) can cancel this
divergent term in the denominators.  The massless limits of the
Proca-like equations are actually
\begin{equation}
\partial_\alpha F^{\alpha\mu} = -{m\over 2} A^\mu
\Longrightarrow \partial_\alpha F^{\alpha\mu} = \partial^\mu \chi\,.
\end{equation}
\end{itemize}

I want to present some comments:
\begin{itemize}
\item
Of course, when we pass over to the second quantization, the commutation
relations for $F^{\mu\nu}$ and $A^\mu$ may be changed in order to keep the
correct dimensions of the fields and in order the action
to be {\it dimensionless}.

\item
Ogievetski\u{\i} and Polubarinov, Kalb and Ramond~\cite{op,kr}
analized the scalar Lagrangian of the antisymmetric tensor
field~\cite{dvoo,dvob,dvohpa} and ``gauge out" the {\it transverse}
components by means of
\begin{equation} F_{\mu\nu} \rightarrow F_{\mu\nu}
+\partial_\nu \Lambda_\mu - \partial_\mu \Lambda_\nu\,,
\end{equation} the
new ``gauge" transformation. Therefore, they obtained a {\it pure}
longitudinal field, the {\it notoph} (or, the {\it Kalb-Ramond field}, as
it is frequently called in the US literature).

\item
In~\cite{dvob,dvohpa} we found a map between the
Ogievetski\u{\i}-Polubarinov formulation and the Weinberg $2(2s+1)$
theory~\cite{weinb}. In the latter case the Lagrangian is given by
\begin{equation}
{\cal L} = \partial_\mu \overline{\Psi}_{(6)} \gamma^{\mu\nu} \partial_\nu
\Psi_{(6)} \pm m^2 \overline{\Psi}_{(6)}\Psi_{(6)}\,,\label{lag}
\end{equation}
(or its analogues for fields of different dimensions).
The $\gamma^{\mu\nu}$ is a  set of covariant matrices of the
$(1,0)\oplus(0,1)$ representation; $\Psi_{(6)}$ and
$\overline{\Psi}_{(6)}$ are bivectors. In general, various $\Psi$ can be
used, which differ each other by discrete symmetry transformations. The map
exists between the equations obtained from (\ref{lag})
\begin{equation}
[\gamma^{\mu\nu} \partial_\mu\partial_\nu \mp m^2 ] \Psi_{(6)} =0
\end{equation}
and the equation
\begin{equation}
\partial_\mu \partial^\alpha
F_\alpha^{\quad \nu} -\partial^\nu \partial^\alpha F_{\alpha\mu} - {1\over 2}
(m^2 + \partial_\lambda \partial^\lambda) F_{\mu\nu} = - m^2
F_\mu^{\quad \nu} \end{equation}
and its dual. See also~\cite{Dva}.

\end{itemize}

Next, we return to the van der Waerden-Sakurai derivation of the Dirac
equation:
\begin{equation}
(E^2 -c^2 \vec{\bf p}^{\,2}) I^{(2)}\Psi_{(2)} =
\left [E I^{(2)} - c {\bm \sigma}\cdot {\bf p} \right ]
\left [E I^{(2)} + c {\bm \sigma}\cdot {\bf p} \right ]
\Psi_{(2)} = m_2^2 c^4\Psi_{(2)}\,.\label{G1}
\end{equation}
If one denotes $\Psi_{(2)}=\eta$ one can define $\chi = {1\over m_1 c}
(i\hbar {\partial\over \partial x_0} - i\hbar {\bm \sigma}\cdot
{\bm\nabla}) \eta$. Please note that
we introduced the second mass parameter $m_1$.
The corresponding set of 2-component equations is
\begin{eqnarray}
&&(i\hbar {\partial \over \partial x_0}
-i\hbar {\bm \sigma}\cdot {\bm \nabla}) \eta
 =m_1 c\chi\,,\\
&&(i\hbar {\partial \over \partial x_0}
+i\hbar {\bm \sigma}\cdot {\bm \nabla}) \chi
={m_2^2 c\over m_1}\eta\,.
\end{eqnarray} 
In the 4-component form  we have
\begin{eqnarray}
&&\pmatrix{i\hbar (\partial/\partial x_0) &
i\hbar {\bm \sigma}\cdot {\bm \nabla}\cr
-i\hbar {\bm \sigma}\cdot {\bm \nabla}&
-i\hbar (\partial/\partial x_0)}
\pmatrix{\chi+\eta\cr\chi-\eta} = \\
&=&{c\over 2}
\pmatrix{(m_2^2/m_1
+m_1)&
(-m_2^2/m_1 +
m_1)\cr
(-m_2^2/m_1 +
m_1)& (m_2^2/m_1
+m_1)\cr}\pmatrix{\chi+\eta\cr\chi-\eta\cr}\,,
\end{eqnarray}
which results in
\begin{equation}
\left [i\hbar \gamma^\mu \partial_\mu - {m_2^2 c \over m_1}
{1-\gamma_5 \over 2} -m_1 c {1+\gamma_5 \over 2}\right ]
\Psi_{(4)} = 0\,.
\end{equation}
The ``new" massless equation is ($m_2\rightarrow 0$, $p^2 =0$, $m_1 \neq 0$)
\begin{equation}
\left [ i\gamma^\mu \partial_\mu - {m_1 c\over \hbar} {1+\gamma_5
\over 2}\right ]\Psi_{(4)} =0\,.\label{nme}
\end{equation}
It is easy to check that dispersion relations are $E=\pm \vert {\bf
p}\vert$, that give us rights to call it {\it massless}, even though
there is a ``mass" parameter in (\ref{nme}). 

In the 2-component formalism we do not know the parity properties of the 2-spinors
$\Psi_{(2)}$. It is possible to obtain yet another equation differing from (\ref{nme})
by the sign at the $\gamma_5$ term. The equation is
\begin{equation} [i\gamma^\mu \partial_\mu - {m_3 c\over
\hbar} {1-\gamma_5 \over 2}]\Psi^\prime_{(4)} = 0 \,.
\end{equation}
Moreover, instead of $(1\pm \gamma^5)/2$ one can use any singular
$4\times 4$ matrix of appropriate physical dimension and still have
massless particles.
The relevant equations can be found in the old~\cite{antec}
and new literature~\cite{Rasp1,Rasp5} literature.

Is the physical content of the generalized $S=1/2$
{\it massless} equations the same as that of the Weyl equation?
Our answer is `No'. The excellent discussion can be found
in~[17a,b]. First of all, the theory does {\it not} have chiral invariance. Those authors
call the additional parameters as measures of the degree of chirality.
Apart of this, Tokuoka introduced the concept of the gauge transformations
(not to confuse with phase transformations) for the 4-spinor fields. He
also found some strange properties of the  anti-commutation relations
(see \S 3 in~[17a]). And finally, the equation describes
{\it four} states, two of which answer for the positive energy $E=\vert
{\bf p}\vert$, and two others answer for the negative energy $E=-\vert
{\bf p}\vert$. I just want to add the following to the discussion.
The operator of the {\it chiral-helicity} $\hat\eta = ({\bm\alpha}\cdot
\hat{\bf p})$  does {\it not} commute, {\it e.g.}, with the Hamiltonian of the
equation~(\ref{nme}):\footnote{Do not confuse with the Dirac Hamiltonian.}
\begin{equation} [{\cal H}, {\bm\alpha}\cdot \hat{\bf
p} ]_- = 2{m_1 c \over \hbar} {1-\gamma^5 \over 2} ({\bm \gamma}\cdot
\hat{\bf p})\, .
\end{equation}
For the eigenstates of the {\it chiral-helicity} the system of corresponding
equations can be read ($\eta=\uparrow, \downarrow$)
\begin{equation} i\gamma^\mu
\partial_\mu \Psi_\eta - {m_1 c\over \hbar}{1+\gamma^5 \over 2}\Psi_{-\eta}
=0 \, .  \end{equation} The conjugated eigenstates of the Hamiltonian
$\vert \Psi_\uparrow + \Psi_\downarrow >$ and
$\vert \Psi_\uparrow - \Psi_\downarrow >$
are connected, in fact, by $\gamma^5$ transformation
$\Psi \rightarrow \gamma^5 \Psi \sim ({\bm\alpha}\cdot \hat{\bf p})
\Psi$ (or $m_1\rightarrow -m_1$).  However, the $\gamma^5$ transformation
is related to the $PT$ ($t\rightarrow - t$ only) transformation~[17b],
which, in its turn, can be interpreted as the change of the energy sign 
$p_0\rightarrow -p_0$, if one
accepts the Stueckelberg idea about antiparticles.  For example, we 
may associate $\vert
\Psi_\uparrow + \Psi_\downarrow >$ with the positive-energy eigenvalue of
the Hamiltonian and $\vert \Psi_\uparrow -
\Psi_\downarrow >$, with the negative-energy eigenvalue of the
Hamiltonian. Thus, the free chiral-helicity
massless eigenstates may oscillate one to another with the frequency
$\omega = E/\hbar$ (as the massive chiral-helicity eigenstates, see~\cite{DVON2}
for details). Moreover, a special kind of interaction which is not
symmetric with respect to the chiral-helicity states (for instance, if
the left chiral-helicity eigenstates interact with the matter only) may induce
changes in the oscillation frequency, like in the Wolfenstein (MSW) formalism.

\section{The Second-Order Equations.}

A correct equation for  an adequate description of  neutrinos
was sought for a long time~\cite{antec,LY,Fush1}. This problem
is, in general, connected with the problem of taking the massless limit of
relativistic equations. For instance, it has been known for a long time
that {\it ``one cannot simply set the mass equal to zero in a manifestly
covariant massive-particle equation, in order to obtain the corresponding
massless case"}, e.~g., Ref.~[17d].

Secondly, in the seventies the second-order equation in the
4-dimensional representation of the $O(4,2)$ group was proposed by Barut
{\it et al.} in order to solve the problem of the mass splitting of
leptons~\cite{Bar0,Wilson,Bar} and by Fushchich {\it et al.}, for
describing various spin states in this representation~\cite{Fush,Fush2}.
The equations (they proposed) may depend on two parameters. Recently we
derived the Barut-Wilson equation on the basis of the first
principles~\cite{DVB}. Briefly, the scheme for derivation of the
equation
\begin{equation}
\left [i\gamma^\mu \partial_\mu  + \alpha_2 \partial^\mu
\partial_\mu  -\kappa \right ] \phi (x)
= 0\,\,   \label{Barut}
\end{equation}
is the following. First, apply the generalized Ryder relation~\cite{Ryder}
(see also below, Eq. (\ref{RB})) and the standard scheme for
the derivation  of relativistic wave equations~\cite[footnote \# 1]{Ahlu},~\cite{Dva}.
Then form the  Dirac 4-spinors; the left- and right parts of them are
connected as follows:
  \begin{eqnarray}
\phi_{_L}^\uparrow (p^\mu) &=& - \Theta_{[1/2]} [\phi_{_R}^{\downarrow}
(p^\mu)]^\ast \quad,\quad \phi_{_L}^\downarrow (p^\mu) = + \Theta_{[1/2]}
[\phi_{_R}^{\uparrow}(p^\mu)]^\ast \,\, ,\label{1a}\\
\phi_{_R}^\uparrow (p^\mu) &=& -
\Theta_{[1/2]} [\phi_{_L}^{\downarrow} (p^\mu)]^\ast \quad,\quad
\phi_{_R}^\downarrow (p^\mu) = + \Theta_{[1/2]}
[\phi_{_L}^{\uparrow} (p^\mu)]^\ast \label{1b}\,\,,
\end{eqnarray} 
in order to obtain
\begin{equation}
\left [a \,{i\gamma^\mu \partial_\mu \over m}
+b\, {\cal C} {\cal K} - 1\right ] \Psi (x^\mu) = 0\label{de}
\end{equation}
in the coordinate space.
Transfer to the Majorana
representation with the unitary matrix
\begin{equation} U ={1\over
2}\pmatrix{1 -i\Theta_{[1/2]} & 1 +i\Theta_{[1/2]}\cr
-1 -i\Theta_{[1/2]} & 1 -i\Theta_{[1/2]}\cr}\quad,\quad
U^\dagger = {1\over 2}\pmatrix{1 -i\Theta_{[1/2]} & -1
-i\Theta_{[1/2]}\cr 1 + i\Theta_{[1/2]} & 1
-i\Theta_{[1/2]}\cr}\,\,.\label{maj}
\end{equation}
Finally, one obtains
the set
\begin{eqnarray}
\left [ a {i\gamma^\mu \partial_\mu \over m} - 1 \right ] \phi - b
\,\chi &=& 0 \,\,,\\
\left [ a {i\gamma^\mu \partial_\mu \over m} - 1 \right ]\chi  - b
\,\phi &=& 0\,\,
\end{eqnarray}
for $\phi (x) = \Psi_1 +\Psi_2$ or $\chi (x)=\Psi_1 -\Psi_2$
(where $\Psi^{^{MR}} (x) = \Psi_1 +i\Psi_2$).
With the identification $a/2m \rightarrow \alpha_2$ and $m(1-b^2)/2a
\rightarrow \kappa$ the above set leads to the second-order equation of
the Barut type.

Thirdly, we found the possibility of generalizations of the $(1,0)\oplus
(0,1)$ equations (namely, the Maxwell's equations and the
Weinberg-Tucker-Hammer equations\footnote{In general, the latter does
{\it not} completely reduce to the former after taking the massless limit
in the accustomed way.}) also on the basis of including two independent
constants~\cite{dvohpa}.

The following definitions and postulates are used in this Section:
The operators of the discrete symmetries are defined as follows:
a) the space inversion operator:
\begin{eqnarray}
P_{[1/2]} = \pmatrix{0&1\cr
1 & 0\cr}
\end{eqnarray}
is the $4\times 4$ anti-diagonal matrix;
b) the charge conjugation operator:
\begin{eqnarray}
C_{[1/2]} = \pmatrix{0 & i\Theta_{[1/2]}\cr
-i\Theta_{[1/2]} & 0\cr}{\cal K}\,, \quad
\Theta_{[1/2]}=\pmatrix{0&-1\cr 1&0\cr}\,,\label{cco}
\end{eqnarray}
with ${\cal K}$ being the operation of complex conjugation.

The left- and the right-  spinors  transform to the frame with
the momentum $p^\mu$ (from the zero-momentum frame) according to the Wigner rules,
with $\Lambda_{_{R,L}} =\exp (\pm {\bf S}\cdot {\bm \varphi})$ being the matrices of the Lorentz boosts. ${\bf S}$ are
the spin matrices for spin $s$, e.~g., Ref.\cite{Var}; ${\bm \varphi}$
are parameters of the given boost.  They are defined, {\it e.~g.}, Refs.~\cite{Ryder,Dva}, by means
of
\begin{equation}\label{boost} \cosh (\varphi) =\gamma =
\frac{1}{\sqrt{1-v^2}} = \frac{E}{m},\quad \sinh (\varphi) = v\gamma =
\frac{\vert {\bf p}\vert}{m},\quad \hat {\bm \varphi} = {\bf n} =
\frac{{\bf p}}{\vert {\bf p}\vert}\,\, .
\end{equation}

The Ryder relation between spinors in the zero-momentum
frame~\cite{Ryder} is established\footnote{It can be derived from the Faustov work~\cite{Faust} too.}
\begin{equation} 
\phi_{_L}^h (\mathring{p}^\mu) = a (-1)^{{1\over 2} - h} e^{i(\vartheta_1
+\vartheta_2)} \Theta_{[1/2]} [\phi_{_L}^{-h} (\mathring{p}^\mu)]^\ast + b
e^{2i\vartheta_h} \Xi^{-1}_{[1/2]} [\phi_{_L}^h (\mathring{p}^\mu)]^\ast
\,\, ,\label{RB}
\end{equation}
$\mathring{p}^\mu =(m,{\bf 0})$
with the {\it real} constant $a$ and $b$ being arbitrary at this stage;
$h$ is the polarization index. Next,
\begin{eqnarray}
\Xi_{[1/2]} = \pmatrix{e^{i\phi}&0\cr
0&e^{-i\phi}\cr}\,\, ,
\end{eqnarray}
$\phi$ is here the azimuthal angle related to ${\bf p} \rightarrow {\bf
0}$.\footnote{In
general, one can connect also $\phi_L^\uparrow$ and $\phi_L^\downarrow$.
with using the $\Omega$ matrix (see formulas (22a,b) in Ref.~\cite{Ahlu}):
\begin{equation}
\phi_L^\uparrow ({0}^\mu) = \Omega \phi_L^\downarrow
({0}^\mu)\,\, , \quad
\Omega =\pmatrix{ cotan (\theta/2) & 0\cr
0& - tan (\theta/2)\cr } = {\vert {\bf p}\vert \over \sqrt{ {\bf p}^{\,2}
- p_3^2} } (\sigma_3 + {p_3 \over \vert {\bf p} \vert })\,\, .
\end{equation}
We did not yet find the explicitly covariant form of the resulting
equation.}

One can form either Dirac 4-spinors:
\begin{equation}
u_h (p^\mu) =\pmatrix{\phi_{_R} (p^\mu)\cr
\phi_{_L} (p^\mu)\cr}\quad,\quad
v_h (p^\mu) = \gamma^5 u_h (p^\mu)\,\, ,
\end{equation}
or the second-type spinors~\cite{Ahlu,BWW}, see also~\cite{Sokol,DVON1,bz}:
\begin{equation}
\lambda (p^\mu) = \pmatrix{(\zeta_\lambda \Theta_{[j]}) \phi_{_L}^\ast
(p^\mu) \cr \phi_{_L} (p^\mu)\cr}\,,\quad
\rho (p^\mu) = \pmatrix{\phi_{_R}
(p^\mu) \cr (\zeta_\rho \Theta_{[j]})^\ast \phi_{_R}^\ast
(p^\mu)\cr}\,\, ,
\end{equation}
or even more general forms of 4-spinors
depending on the phase factors between their left- and right- parts and
helicity sub-spaces that they belong to.
For the second-type spinors
several forms of the field operators have been proposed. For example,
\begin{eqnarray}
\nu^{DL} (x^\mu) &=& \sum_\eta \int \frac{d^3 {\bf p}}{(2\pi)^3}
{1\over 2E_p} \left [ \lambda^S_\eta (p^\mu) c_\eta (p^\mu) \exp (-ip\cdot
x) +\right.\nonumber\\
&+&\left.\lambda^A_\eta (p^\mu) d_\eta^\dagger (p^\mu) \exp
(+ip\cdot x)\right ]\,\, .
\end{eqnarray}

The Dirac equation has been derived by this method (the relation between  2-spinors at rest. 
$\phi_R ({\bf 0}) = \pm \phi_L ({\bf 0})$, and boosts). Next, the coupled Dirac equations for $\lambda -$ and $\rho -$ 
spinors have also been presented~\cite{DVON2,DVON1}. The corresponding Lagrangian, 
projection operators, and the Feynman-Dyson-Stueckelberg propagator have been found later.
However, we have surprisingly:
\begin{eqnarray}
d^\dagger_\kappa (p)&=& -\frac{ip_y}{p}\sigma^y_{\kappa\tau}  c_\tau (-p)\,,\\
c_\kappa (-p)&=& -\frac{ip_y}{p}\sigma^y_{\kappa\tau}  d^\dagger_\tau (p)\,.
\end{eqnarray}
In the Majorana-like case ($c_\eta (p)= e^{-i\varphi} d_\eta (p)$) we have difficulties in the construction of field operators.


On the basis of these definitions on using the standard
rules~\cite[footnote \# 1]{Ahlu}  one can  derive:
In the case $\vartheta_1 =0$, $\vartheta_2 =\pi$ the following
equations are obtained for $\phi_{_L} (p^\mu)$ and
$\chi_{_R} = \zeta_\lambda \Theta_{[1/2]} \phi_{_L}^\ast (p^\mu)$:\footnote{The 
phase factors $\zeta$ are defined by various constraints
imposed on the 4-spinors, e.~g.,
the condition of the self/anti-self charge conjugacy gives
$\zeta_\lambda^{S,A}=\pm i$. But, one should still note that
phase factors also depend on the phase factor in the definition
of the charge conjugation operator (\ref{cco}). The ``mass term" of
resulting dynamical equations may also be different.}
\begin{eqnarray}
&&\phi_{_L}^h (p^\mu ) = \Lambda_{_L} (p^\mu \leftarrow \mathring{p}^\mu)
\phi_{_L}^h ({0}^\mu)
= {a \over \zeta_\lambda}
(-1)^{{1\over 2} +h} \Lambda_{_L} (p^\mu \leftarrow \mathring{p}^\mu)
\Lambda_{_R}^{-1} (p^\mu \leftarrow \mathring{p}^\mu) \chi_{_R}^h (p^\mu)+
\nonumber\\
&&+{b\over \zeta_\lambda} \Lambda_{_L} (p^\mu \leftarrow {0}^\mu)
\Xi^{-1}_{[1/2]} \Theta^{-1}_{[1/2]} \Lambda_{_R}^{-1} (p^\mu \leftarrow
{0}^\mu) \chi_{_R}^{-h} (p^\mu)\,\,,\\
&&\chi_{_R}^{-h} (p^\mu) = \Lambda_{_R}
(p^\mu \leftarrow {0}^\mu) \chi_{_R}^{-h}
({0}^\mu) =
a \zeta_\lambda (-1)^{{1\over 2}
-h} \Lambda_{_R} (p^\mu \leftarrow {0}^\mu) \Lambda_{_L}^{-1}
(p^\mu \leftarrow {0}^\mu) \phi_{_L}^{-h} (p^\mu) +\nonumber\\
&&+ b\zeta_\lambda \Lambda_{_R} (p^\mu \leftarrow {0}^\mu)
\Theta_{[1/2]} \Xi_{[1/2]} \Lambda_{_L}^{-1} (p^\mu \leftarrow
{0}^\mu) \phi_{_L}^{h} (p^\mu)\,\,.
\end{eqnarray} 
Hence, the equations for the 4-spinors
$\lambda^{S,A}_\eta (p^\mu)$
take the forms:
\begin{eqnarray}
ia
{\widehat p \over m} \lambda^S_\uparrow (p^\mu) - (b{\cal C}{\cal K}
-1) \lambda^S_\downarrow (p^\mu) &=& 0\,\, ,
\label{m1}\\
ia {\widehat p \over m} \lambda^S_\downarrow (p^\mu) + (b
{\cal C}{\cal K} -1) \lambda^S_\uparrow (p^\mu) &=& 0\,\, ,
\label{m2}\\
ia {\widehat p \over m} \lambda^A_\uparrow (p^\mu) - (b
{\cal C}{\cal K} +1) \lambda^A_\downarrow (p^\mu) &=& 0\,\, ,
\label{m3}\\
ia {\widehat p \over m} \lambda^A_\downarrow (p^\mu) + (b {\cal
C}{\cal K} +1) \lambda^A_\uparrow (p^\mu) &=& 0\,\,
\label{m4},
\end{eqnarray}  
$a =\pm (b-1)$ if we want to have $E_p^2 - {\bf p}^2 = m^2$ for massive
particles.
We can write several forms of equations in the coordinate representation
depending on the relations between creation/annihilation operators.
For example,
provided that we imply $d_\uparrow (p^\mu) = +ic_\downarrow (p^\mu)$
and $d_\downarrow (p^\mu) = -ic_\uparrow (p^\mu)$; the ${\cal K}$
operator acts on $q-$ numbers as hermitian conjugation, then
the first generalized equation in the coordinate space reads
\begin{equation}
\left [ ia {\gamma^\mu \partial_\mu \over m} - (b-1) \gamma^5 {\cal
C} {\cal K} \right ] \Psi (x^\mu) = 0\,\, .
\end{equation}
Transferring into the Majorana representation one obtains two
real equations:\footnote{This procedure can be carried out
for any spin, cf.~\cite{Dvo2-1997}.}
\begin{eqnarray}
ia {\gamma^\mu \partial_\mu \over m} \Psi_1 (x^\mu) -i (b-1)
\gamma^5 \Psi_2 (x^\mu) &=& 0\,\, ,\\
ia {\gamma^\mu \partial_\mu \over
m} \Psi_2 (x^\mu) -i (b-1)\gamma^5 \Psi_1 (x^\mu) &=& 0\,\, .
\end{eqnarray}  
for real and imaginary parts of the field
function $\Psi^{^{MR}} (x^\mu) = \Psi_1 (x^\mu) +i\Psi_2 (x^\mu)$.
In the case of $a =1-b$
and considering the field function $\phi= \Psi_1 +\Psi_2$
we come to the equation
for the spinors of the {\it second
kind}~\cite[Eq.(8)]{Sokol} and Ref.~\cite{bz}.
Next, we  come to the  second-order equation in the coordinate
representation for massive particles
\begin{equation} \left [ a^2
{\partial_\mu \partial^\mu \over m^2} +(b-1)^2 \right ] \cases{\Psi_1
(x^\mu) &\cr \Psi_2 (x^\mu) &\cr} = 0\,\,.\label{kg2} \end{equation} Of
course, it may be reduced to the Klein-Gordon equation.
In general, there may exist mass splitting between various $CP-$
conjugate states.

One can find the relation between creation/annihilation
operators for another equation ($\beta_1 , \, \beta_2 \in \Re e$)
\begin{equation}
\left [ ia {\gamma^\mu \partial_\mu \over m} - e^{i\alpha_1}\beta_1
\gamma^5 {\cal C} {\cal K}  + e^{i\alpha_2} \beta_2 \right ] \Psi (x^\mu)
= 0\,\, ,\label{neq1}
\end{equation}
which would be consistent with the equations
(\ref{m1}-\ref{m4}).\footnote{As one can expect from  this consideration
the equation (\ref{neq1}) may be reminiscent of the old works,
Refs.~\cite{antec,Rasp1}.}
In the Majorana representation the resulting set of the real equations
are 
\begin{eqnarray}
&&\left [ia {\gamma^\mu \partial_\mu \over m} +i\beta_1 \sin\alpha_1
\gamma^5  +\beta_2 \right ] \Psi_1 -i\beta_1 \cos\alpha_1 \gamma^5 \Psi_2
= 0\,\, ,\\
&&\left [ia {\gamma^\mu \partial_\mu \over m} -i\beta_1 \sin\alpha_1
\gamma^5  +\beta_2 \right ] \Psi_2 -i\beta_1 \cos\alpha_1 \gamma^5 \Psi_1
= 0\,\, .
\end{eqnarray}  
For instance in the $\alpha_1 = {\pi \over 2}$ we obtain
\begin{eqnarray}
\left [ ia {\gamma^\mu \partial_\mu \over m} +i \beta_1 \gamma^5 +\beta_2
\right ] \Psi_1 &=& 0\,\, ,\label{gf1}\\
\left [ ia {\gamma^\mu \partial_\mu \over m} -i \beta_1 \gamma^5 +\beta_2
\right ] \Psi_2 &=& 0\,\, \label{gf2}.
\end{eqnarray} 
But, in any case one can recover the
Klein-Gordon equation for both real and imaginary parts of the field
function, Eq. (\ref{kg2}).  

We are able to consider other constraints on the
creation/annihilation operators, introduce various types of
fields operators (as in~\cite{dvohpa}) and/or generalize the
Ryder relation even more. In the general case, we suggest to start from
\begin{equation}
(a{\widehat p \over m} -1) u_h (p^\mu) +i b (-1)^{{1\over 2}-h}
\gamma^5 {\cal C} u_{-h}^\ast (p^\mu) =0\,\,;
\end{equation}
i.~e., the equation (11) of~\cite{DVB}. But, as opposed to the cited paper,
we write the coordinate-space equation in the form:
\begin{equation}
\left [a \,{i\gamma^\mu \partial_\mu \over m}
+b_1\, {\cal C} {\cal K} - 1\right ] \Psi (x^\mu)
+ b_2 \gamma^5  {\cal C} {\cal K} \widetilde{\Psi} (x^\mu)
= 0\,\, ,\label{dem}
\end{equation}
thus introducing the third parameter. Then we can perform
the same procedure as in Ref.~\cite{DVB}. Implying $\Psi^{^{MR}}
=\Psi_1 + i\Psi_2$ and $\widetilde \Psi^{^{MR}}
=\Psi_3 + i \Psi_4$, one obtains real equations in the Majorana
representation:
\begin{eqnarray}
(a {i\gamma^\mu \partial_\mu \over m} - 1 )\phi
-b_1 \chi +ib_2 \gamma^5 \widetilde \phi &=& 0\, ,\\
(a {i\gamma^\mu \partial_\mu \over m} - 1 )\chi
-b_1 \phi -ib_2 \gamma^5 \widetilde \chi &=& 0\, ,
\end{eqnarray}
for $\phi = \Psi_1 +\Psi_2$, $\chi =\Psi_1 - \Psi_2$
and $\widetilde\phi = \Psi_3 +\Psi_4$, $\widetilde\chi =\Psi_3 - \Psi_4$.
After algebraic transformations we have:
\begin{eqnarray}
&& (a{i\gamma^\mu \partial_\mu \over m} - b_1 -1 ) \left [ 2ia
{\gamma^\nu \partial_\nu \over m} + a^2 {\partial^\nu \partial_\nu \over
m^2} + b_1^2 - 1 \right ] \Psi_1  -\nonumber\\
&& \qquad - ib_2 \gamma^5 \left [ 2ia
{\gamma^\mu \partial_\nu \over m} - a^2 {\partial^\mu \partial_\mu \over
m^2} - b_1^2 + 1 \right ] \Psi_4 = 0\, ;\\
&& (a{i\gamma^\mu \partial_\mu \over m} + b_1 -1 ) \left [ 2ia
{\gamma^\nu \partial_\nu \over m} + a^2 {\partial^\nu \partial_\nu \over
m^2} + b_1^2 - 1 \right ] \Psi_2  -\nonumber\\
&& \qquad - ib_2 \gamma^5 \left [ 2ia
{\gamma^\mu \partial_\mu \over m} - a^2 {\partial^\mu \partial_\mu \over
m^2} - b_1^2 + 1 \right ] \Psi_3 = 0\, ,
\end{eqnarray}
the third-order equations. However, the field operator
$\widetilde \Psi$ may be linear dependent on the states
included in the $\Psi$. So, relations may exist
between $\Psi_{3,4}$ and $\Psi_{1,2}$. If we apply
the simplest constraints
$\Psi_1 = -i\gamma^5 \Psi_4$ and $\Psi_2 =i\gamma^5 \Psi_3$
one should recover the Dirac-Barut-like equation
with {\it three} mass eigenvalues:
\begin{equation}
\left [ i\gamma^\mu \partial_\mu - m {1\pm b_1 \pm b_2 \over a}
\right]\times
\left [ i\gamma^\nu \partial_\nu + {a\over 2m} \partial^\nu \partial_\nu
+ m {b_1^2 -1 \over 2a}\right ] \Psi_{1,2} =0\, .
\end{equation}
Furthermore, we apparently note that the similar results
can be obtained by consecutive applications of the generalized
Ryder relations. 
As indicated by Barut himself, several ways for introdcution of
interaction  with 4-vector potential exist in second-order equations.
Only considering the correct one (and, probably, taking into account
$\gamma^5$ axial currents), we shall be able to
answer the question of  why the $\alpha_2$ parameter of the Barut works is
fixed by means of the use of the {\it classical} value of the anomalous 
magentic moment; and on what physical basis we have to fix other 
parameters we introduced above.

Concerning the massless limit please note that the mass appears in the denominator within
this method. So, we multiply by it in order to obtain the Dirac-like equations that is not an obvious procedure.
Next, we have the d'Alembert operator $\partial_\mu \partial^\mu$, whose eigenvalues may be different from $m$,
and even be zeros. Anyway, a) in every equations we should calculate the characteristic determinant to find out 
the dispersion relations; b) we should not forget about the possibility of divergent (in $m\rightarrow 0$) terms
in the solutions of the corresponding equations, and treat them properly.

\section{Conclusions.}

In conclusion, we presented two very natural ways of deriving the
massive/massless equations in the $(S,0)\oplus (0,S)$ representation
space, which lead to the equations given by other researchers in the past.
It is known that present-day neutrino physics has come across
serious difficulties. Experiments and observations are not in
agreement with theoretical predictions of the standard model.
That was a motivation for the present work. 

The Barut's way of solving the hierarchy problem, which
was almost forgotten, has been analized here from different viewpoint.

The terms $m(1\pm \gamma_5)/2$ in the massless equation violate chiral
invariance of the theory (due to $[\hat p,\gamma_5]_- \neq 0$).
The solutions are {\it not} eigenstates of $\gamma_5$ operator, but in the
case $m_{1,3} \rightarrow 0$ the chiral invariance is restored. So, they
have been called {\it measures of chirality}.

In fact, the solutions 
of the Dirac massless/massive equation 
represent a mixture of various polarization states. This fact may be related to
recent research of the Majorana-like constructs ($m\neq 0$) when we also
mixed solutions of the Dirac equation in order to obtain self/anti-self
charge conjugate states~\cite{ZIINO,bz}.

Dynamics of massless particles and neutrino may differ from those derived from
the well-known Weyl and Maxwell equations.

\medskip

{\bf Acknowledgements.} I greatly appreciate discussions with Profs. A. Gersten, A. Raspini and  A. F. Pashkov. 
This work has been partly supported by the Mexican Sistema Nacional de Investigadores and the ESDEPED. 

\end{document}